\documentclass[pra,aps,twocolumn]{revtex4}
	
\usepackage[dvips]{graphicx}
\usepackage{amsfonts}
\usepackage{amscd}

\newtheorem{theorem}{Theorem}

\newtheorem{lemma}[theorem]{Lemma}

\newtheorem{proposition}[theorem]{Proposition}

\newtheorem{claim}[theorem]{Claim}
\newcommand{\qed}{\rule{7pt}{7pt}}
\newenvironment{proof}{\noindent{\bf Proof}\hspace*{1em}}{\qed\bigskip}

\newcommand{\bra}[1]{\mbox{$\langle #1 |$}}
\newcommand{\ket}[1]{\mbox{$| #1 \rangle$}}
\newcommand{\braket}[2]{\mbox{$\langle #1 | #2 \rangle$}}
\def\tr{\mbox{Tr}\,}
\def\be{\begin{equation}}
\def\ee{\end{equation}}
\def\bea{\begin{eqnarray}}
\def\eea{\end{eqnarray}}
\def\ben{\begin{eqnarray*}}
\def\een{\end{eqnarray*}}
\def\n{\nonumber}
\def\B{\mathcal{B}}
\def\H{\mathcal{H}}
\def\sig{\mathcal{S}}
\def\rank{\mbox{rank}\,}

\begin{document}

\title{A tight lower bound on the classical communication cost
of entanglement dilution}

\author{Aram W. Harrow}
\email{aram@mit.edu}
\affiliation{MIT Physics Dept., 77 Massachusetts
Avenue, Cambridge, MA 02139}
\author{Hoi-Kwong Lo}
\affiliation{MagiQ Technologies, Inc.,
275 Seventh Avenue, 26th floor,
New York, NY 10001-6708, USA}

\date{\today}

\begin{abstract}
In this paper we investigate the classical communication cost of
converting between different forms of bipartite pure state
entanglement in the many copy case.  This problem is usually broken
into two parts: {\em concentrating} the entanglement from many
partially entangled states into a smaller number of singlets and the
reverse process of {\em diluting} singlets into
partially entangled states.

Entanglement concentration requires no classical communication, but
the best prior art result for diluting to $N$ copies of a partially
entangled state requires an amount of communication on the order of
$\sqrt{N}$.  Our main result is to prove that this prior art result is
optimal up to a constant factor; any procedure for approximately
creating $N$ partially entangled states from singlets requires
$\Omega(\sqrt{N})$ bits of classical communication.  Previously not
even a constant bound was known for approximate entanglement
transforms.

We also prove a lower bound on the inefficiency of the process: to
dilute singlets to $N$ copies of a partially entangled state, the
entropy of entanglement must decrease by $\Omega(\sqrt{N})$.

\vspace*{4ex}
\end{abstract}

\maketitle

\section{Background}

A useful paradigm in quantum information processing is the resource
model where various entities including classical communication cost,
quantum communication cost and entanglement are regarded as different
fundamental resources.  For a resource model to make sense, different
forms of the same resource
need to be inter-convertible or fungible in the asymptotic limit.  For
example, many independent copies of any classical random variable with
entropy $H$ can be compressed asymptotically losslessly to $H$ bits
per copy \cite{CT}.  Similarly, a quantum state $\rho$ can be
compressed to $S(\rho)$ qubits in the many copy
limit \cite{sch,jozsa}.

In this paper, we will discuss the resource of entanglement:
specifically, bipartite pure state entanglement.  This sort of entanglement
was proposed as asymptotically fungible (and thus a resource) by
Bennett, Bernstein, Popescu and Schumacher (BBPS) \cite{Bennett95}.
If $\ket{\psi}_{AB}$ is a bipartite pure state with entropy of
entanglement $E = S(\tr_B\ket{\psi}\bra{\psi})$ and $\ket{\Phi}$ is
the singlet state $\frac{1}{\sqrt{2}}(\ket{01}-\ket{10})$, BBPS
explained how to approximately convert $\ket{\Phi}^{nE}$ into
$\ket{\psi}^n$ and back again, using local operations and classical
communication (LOCC) only.  The transformations, known as
entanglement concentration and dilution, are only asymptotically
reversible; in each direction we accept some inefficiency (so that
instead of converting $\ket{\Phi}^{nE}$ to $\ket{\psi}^n$, we need to
start with $\ket{\Phi}^{nE+o(n)}$) and a small error $\epsilon\in
o(1)$ (so instead of obtaining the state $\ket{\psi}^n$,
we get $\rho$ with $\tr |\rho - \ket{\psi} \bra{\psi}^n| <
\epsilon$).

An additional feature of the BBPS protocol for entanglement dilution
is a classical communication cost of $O(n)$ bits.  In contrast, their
entanglement concentration procedure requires no communication at all.
For some applications such as super-dense coding \cite{BW92}, paying
an $O(n)$ classical communication cost would mean that the utility of
entanglement for different tasks depended on its form.  If there were
scenarios in which different forms of pure entanglement were not
asymptotically equivalent, then the resource model of entanglement
could become much more complicated.  Fortunately, in 1999 Lo and
Popescu \cite{LP99} showed that the original dilution protocol could
be modified to require only $o(n)$ bits of classical communication,
while the error and inefficiency remained $o(1)$ and $o(n)$
respectively.  Thus, any two states with the same entropy of
entanglement are asymptotically interconvertible, even if we take into
account the cost of classical communication.  The specific dilution
procedure of \cite{LP99} used $O(\sqrt{n})$ bits, but left open the
question of whether this could be improved.  The main result of this
paper proves that no such improvement is possible. In other words, we
will show that $\Omega (\sqrt{n})$ bits of classical communication are
necessary for entanglement dilution.  Previously it was known that
some exact entanglement transformations were only possible with a
non-zero amount of classical communication, but for approximate
transformations no bounds were known.  We also prove a similar lower
bound on inefficiency; creating $n$ states, each of entanglement $E$,
requires starting with $nE+\Omega(\sqrt{n})$ singlets to achieve
$O(1)$ error.

A main motivation of our study is to understand fluctuations in a
finite system.  For entanglement manipulation, these take the form of
probability of failure, imperfect fidelity, suboptimal yield and
required classical communication. Given that any physical system must
have only a finite number of constituents, it is
important to understand the magnitude and origin of those problems.
That is to say: How quickly does a finite-copy system approach its
thermodynamic/asymptotic limit as its number of copies increases?  As
we will see later, some of those problems are related to the fact that
when different states are not related to each other in trivial ways,
we can only approximate them.  Others are related to the spectrum of
the Schmidt coefficients.

Errors and the probability of failure are results of discretization
and can be made exponentially small in $n$.  In contrast, we will show that
the inefficiency and the classical communication cost of entanglement
dilution are necessarily $\Omega(\sqrt{n})$ because they stem from
differences in the shapes of spectra of the Schmidt coefficients of
different states.  We will make this notion precise later, by defining
the {\it significant subspace} (or $\delta$-significant subspace) of a
density matrix $\rho$ to be a subspace that contains at least $\delta$
of the weight of $\rho$ for some $O(1)$ constant
$\delta$ 
\footnote{The idea of a $\delta$-significant subspace has
a classical analog. In Chapter Three of \cite{CT},  the idea of
a high probability set for a distribution is described. We
thank Debbie Leung for bringing this point to our attention.}.
This definition generalizes typical subspaces, which are usually
defined as containing almost all the weight of a density matrix. We
will show that, for a generic $\rho$ that is neither pure nor
maximally mixed, the ratio of the size of a typical subspace of
$\rho^n$ to that of a $\delta$-significant subspace (say with $\delta
= 1/4$) of $\rho^n$ is large, namely $2^{ \Omega (\sqrt{n}) }$.  It is
the logarithm of this ratio that gives rise to the fundamental
constraint---that the inefficiency and classical communication cost of
entanglement dilution have lower bounds of $\Omega(\sqrt{n})$.

To put these results in perspective, it is worth noting that the
entanglement dilution and concentration protocols of BBPS both reduce
the entropy of entanglement by $O(\sqrt{n})$.  In both cases, this
amount of inefficiency turns out to be optimal.  We will prove the
dilution bound in section~\ref{sec:main-proof}, and for concentration,
\cite{LP97} proved that $o(1)$ error requires an inefficiency of
$\Omega(\sqrt{n})$
\footnote{This bound applies only to concentration protocols that
create a deterministic number of singlets.  If we relax this
assumption, then $n$ states with entanglement $E$ can be converted to
$m$ singlets where $m$ is a random variable with expectation
$nE-O(\log n)$ \cite{KM01}. For dilution, however, our
$\Omega(\sqrt{n})$ bound applies equally well to variable yield
dilution protocols since the proof is highly insensitive to the
protocol's success probability.}. 
 Similarly, both classical and quantum data compression require
$\Omega(\sqrt{n})$ more space (either bits or qubits) to compress a
source than would be implied by the entropy of the source
 \footnote{This well-known result is a consequence of our
Proposition~\ref{prop:binomial-significant}.}.

This paper is organized as follows.  Section~\ref{sec:main-statement}
contains the formal statement of our main result---that entanglement
dilution necessarily requires $\Omega(\sqrt{n})$ classical bits of
communication and an inefficiency of $\Omega(\sqrt{n})$.  In the next
two sections, we present useful intermediate results:  In
Section~\ref{sec:protocol-reduction}, we show that a general strategy
of entanglement dilution can be equivalently rephrased as a much
simpler one.  In Section~\ref{sec:sig-subspace} we define
$\delta$-significant subspaces and study their properties.  Our main
result is proved in Section~\ref{sec:main-proof}, and
Section~\ref{sec:conclusion} contains some concluding remarks and
discussions.

\section{Statement of the main result}\label{sec:main-statement}

If a partially entangled bipartite pure state
$\ket{\psi}$ has entropy of entanglement $E$,
then $\ket{\psi}^n$ can be approximately prepared by two distant
parties, Alice and Bob, from roughly $nE$ singlets using only LOCC.
The main result of the present paper is that any such dilution
procedure must use $\Omega (\sqrt{n})$ bits of classical
communication.  Along the way, we will also prove that dilution
protocols cannot be perfectly efficient, and inevitably waste $\Omega
(\sqrt{n})$ bits of entanglement.  The formal statement of our main
result is as follows:

\begin{theorem}\label{thm:main-result}
Let $\ket{\psi}\in\H_{AB}$ be a bipartite pure state that is neither
separable nor maximally entangled with its entropy of entanglement
$E=S\left(\tr_{B}\ket{\psi}\bra{\psi}\right)$ where
$S\left(\tr_{B}\ket{\psi}\bra{\psi}\right)$ is the entropy of a
reduced density matrix.  Let
$\ket{\Phi_d}=\frac{1}{\sqrt{d}}\sum_{i=1}^d \ket{i}\ket{i}$ be a
maximally entangled state of dimension $d$.  Then there exist a
universal constant $\epsilon_0$ and constants $\alpha$ and $n_0$ that
depend on $\ket{\psi}$ such that $\forall\epsilon\leq\epsilon_0$,
$\forall n\geq n_0$, any entanglement dilution protocol transforming
$\ket{\Phi_d}$ into $\ket{\psi}^n$ with error $\epsilon$, probability
of success $2^{-s}$ and using $c$ bits of classical communication must
have
\begin{itemize}
\item[a)]
$\log d\geq nE + \alpha \sqrt{n}$ and
\item[b)]
$c+s\geq\alpha\sqrt{n}$.
\end{itemize}
\end{theorem}

This establishes the lower bounds as strongly as possible, by
requiring only a constant bound on the error and disallowing the
possibility of a trade-off between inefficiency and
classical communication cost. In the above Theorem, we have made use
of the following definitions:

{\em Definition of the trace distance}: The trace distance is defined
as
\be D(\rho,\sigma)=\tr|\rho-\sigma|. \label{eq:trace-def}\ee
We will often use the equivalent formulation
\be D(\rho,\sigma)=2\max_P\tr(P(\rho-\sigma))\label{eq:trace-dist}\ee
where $P$ is a projector \cite{Fuchs95}.  It is
important not to confuse this distance with the matrix norm, which we
define in the usual way.

{\em Definition of the norm of a matrix:} By the norm of a matrix $A$,
denoted $\|A\|$, we mean its largest singular value.  Equivalently,
$\|A\|=\sup_{|v|=1} |Av|$.

{\em Definition of error:} By error $\epsilon$, we
mean that upon success the protocol outputs not $\ket{\psi}^n$, but some
possibly mixed state $\sigma$ with $D(\sigma,\ket{\psi}^n)=\epsilon$.  For
this proof we require only that the protocol has an error
$\epsilon$ smaller
than a universal constant $\epsilon_0\approx 0.01$.

{\em Definition of failure:} Failure, on the other hand, means that
sometimes Alice's measurement yields a state that is far from
$\ket{\psi}^n$.  

{\em Remark:} Our lower bound on classical communication,
stated in Part b) of Theorem~\ref{thm:main-result}, still holds
even if the probability of a protocol succeeding is vanishingly small
(i.e. $2^{-o(\sqrt{n})}$).  At first glance this might seem to be a
surprisingly strong result.  Why should our bound apply so broadly to
probabilistic protocols?  When we prove Theorem~\ref{thm:main-result}
(in Section~\ref{sec:main-proof}) we will find that the communication
bound is independent of the amount of prior entanglement used.  As a
result, it is possible to convert any probabilistic protocol into a
nearly deterministic protocol and vice versa in the following manner:

Suppose there exists a protocol with a $2^{-s}$ probability of success
(and $c$ bits of classical communication).  We will show that it is
always possible to perform a modified protocol with $1-\epsilon$
probability of success and $c+s+O(\log\log(1/\epsilon))$ bits of
communication, at the cost of massively increasing the inefficiency
(i.e., the loss of entanglement).
To implement the modified protocol, Alice performs her measurement
$2^sO(\log 1/\epsilon)$ times on different inputs and with $\approx
1-\epsilon$ probability she will succeed on at least one of them.  She
can then transmit the index of the successful block in
$s+O(\log\log(1/\epsilon))$ bits and send the $c$-bit measurement
result corresponding only to the successful outcome while discarding
the failures.

Conversely, $c$ bits of classical communication can always be
eliminated by having Bob guess the message Alice would have sent with
a $2^{-c}$ probability of success. (In this way, Alice needs to send
Bob at most one additional bit to inform him whether his guess is
correct.)  Thus it is in general impossible to bound either success
probability or classical communication cost independently.  Instead,
Part b) of Theorem~\ref{thm:main-result} gives a tradeoff between
success probability and classical communication cost.

\section{Reducing entanglement manipulation protocols to a standard
form}\label{sec:protocol-reduction} A general strategy for
entanglement manipulation may involve two-way classical communication
between Alice and Bob.  Suppose Alice and Bob have some method of
performing an approximate entanglement manipulation procedure using
local operations and $c$ total bits of classical communication (in
either direction).  In entanglement dilution, for example, Alice and
Bob begin sharing some number of perfect singlets and with high
probability end with a mixed state that approximates many copies of a
partially entangled state.  In this section, we will simplify the
description of any such entanglement manipulation procedure.

\begin{claim}\label{claim:protocol-reduction}
Given a pure bipartite initial state and any LOCC entanglement
manipulation protocol by Alice and Bob that involves no more than $c$
bits of classical communication, there exists an equivalent strategy
(meaning it uses the same amount of communication, takes the same
inputs and has the same output distribution) consisting of the
following:
\begin{enumerate}
\item Alice performs a generalized measurement $\{M_k\}$ with no more
than $2^c$ outcomes on her half of the input.
\item She transmits the result to Bob using $c$ bits of classical
communication.
\item Bob performs a unitary operation $U_k$ conditioned on the
result.
\item Both sides discard ancillary systems.
\end{enumerate}
\end{claim}

A similar claim was proved in \cite{LP97}: it was shown that any LOCC
entanglement manipulation strategy that begins and ends with pure
states can be reduced to one that uses only one-way communication.

However, we need to address two new subtleties here.  First, we need
to consider approximate entanglement transformations which begin in
pure states but can end in mixed states.  Second, \cite{LP97} does not
specifically address the issue of classical communication cost.
In our modified protocol we need Alice to transmit her entire
measurement outcome using no more classical communication than the
original protocol.

\begin{proof}
Local operations can be broken into unitary transforms, measurements,
adding ancilla systems, and discarding ancillas \cite{Vidal98}.
Without loss of generality we can add all the ancilla systems at the
beginning of the protocol and discard all the subsystems at the end.

Simplifying measurements and classical communication is more
complicated.  Consider any measurement performed in a dilution
protocol.  By Neumark's theorem \cite{peres}, we can convert such a
measurement into a unitary operation followed by a projective
measurement in the computational basis.  This can be thought of as
coupling the system to a measurement apparatus and then recording the
state of the measuring device.  Denote the effects of a projective
measurement (on a single qubit) by a superoperator, $\$$.
It has Kraus operators
$\ket{0}\bra{0}$ and $\ket{1}\bra{1}$.  We can also represent the
classical channel between Alice and Bob as a quantum channel that maps
input $\rho$ to output $\$(\rho)$.

At this point, the most general protocol is equivalent to one in which
Alice performs a unitary operation, applies $\$$ some number of times,
and transmits some bits via the classical channel; then Bob performs
some unitary operation conditioned on the message, applies $\$$ and
sends some qubits through the classical channel; and so on.
Conditioning a unitary operation on a measurement outcome $M$ can be
written as a single unitary matrix
\be \ket{0}\bra{0} \otimes U + \ket{1}\bra{1}\otimes V 
\label{eq:classical-control-gate}\ee
where $U$ and $V$ are unitary gates operating on the target system.
Since we never rewrite the qubits storing measurement outcomes and
only use them for classical control of the rest of the system, the
only gates we will apply to them will be of the form
in Eq.~(\ref{eq:classical-control-gate}). From
Figure~\ref{fig:locc-commute},
it is easy to see that these commute with $\$$.

\begin{figure}[htbp]
\includegraphics[keepaspectratio,width=3in]{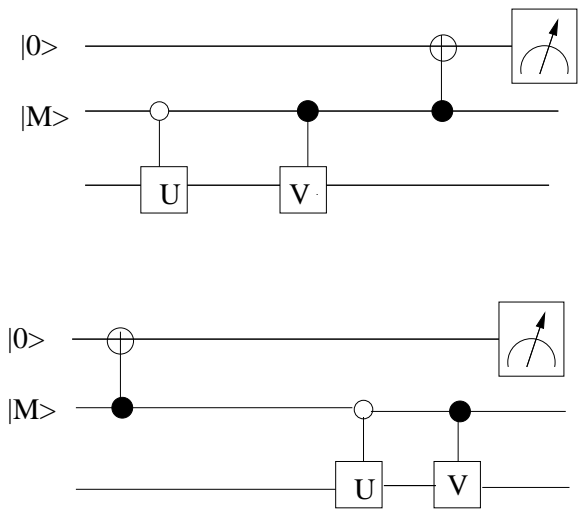}
\caption{Demonstration that $\$$ commutes with the gate in
Eq.~(\ref{eq:classical-control-gate}).  The top line is an ancilla used
to illustrate how $\$$ can be performed, the middle line holds the
measurement outcome $\ket{M}$ and the bottom line represents the rest
of the system which we act on.  Since the two circuits are equivalent,
it follows that $\$$ commutes with any gates of the form in
Eq.~(\ref{eq:classical-control-gate}).}
\label{fig:locc-commute}
\end{figure}

Thus, without loss of generality we can defer every application of
$\$$ until the end of the protocol.

This seemingly trivial step turns out to be equivalent to requiring
Alice and Bob to communicate their complete measurement outcomes.  To
see this, note that until the final discarding of ancillas, $\$$ is
{\it only} applied to the bits that Alice and Bob actually send to
each other.  Equivalently, Alice and Bob only perform projective
measurements on qubits and they always report their measurement
outcomes.  This means that both parties always know the joint state
exactly until the final step when they discard ancillas.  For this
reason, we can deal with pure states only, a situation sometimes
referred as the Church of the larger Hilbert space \footnote{The
Church of the Larger Hilbert space is a fruitful idea in quantum
information theory. It has been applied in, for example, the proof of
the impossibility of quantum bit
commitment \cite{mayersbit,lochaubit}
and quantum
oblivious transfer \cite{Lo97}.
For a review, see, for
example, \cite{CL98}}.

Under these conditions, \cite{LP97} showed that any measurement
performed by Bob where he communicates the outcome to Alice can be
simulated by a measurement by Alice where she communicates the outcome
to Bob (without changing classical communication cost). This result
holds because the Schmidt decomposition of a pure state is always
symmetric under the interchange of Alice and Bob.  Therefore, without
loss of generality, we can consider a reduced protocol where only
Alice performs measurements and at the end communicates the entire
measurement record to Bob. 

As a final simplification, we note that Alice can combine a
whole sequence of measurements into a single measurement.

In summary, given any fixed protocol for
entanglement manipulation, we can construct an equivalent reduced
protocol that consists of the following steps: a generalized
measurement by Alice, transmission of the complete measurement outcome
to Bob, a unitary operation by Bob conditioned on the measurement and
then discarding of ancillas on both sides.  Moreover the modified
protocol uses the same amount of classical communication as the
original protocol.
\end{proof}

\section{Significant subspaces of a density matrix}
\label{sec:sig-subspace}
A {\em typical subspace} of a density matrix $\rho$ is a vector space
that contains most (i.e. $1-o(1)$) of the weight of $\rho$ but in
general has a dimension much smaller than the rank of $\rho$ \cite{CT}.
For our proof we will introduce the related concept of a {\em
significant subspace}, which is a vector space that contains a
significant portion (meaning some $O(1)$ constant) of the weight of
$\rho$.  Neither significant subspaces nor typical subspaces have
properties that are unique to quantum information theory, but we will
find it more convenient to state our definitions in terms of density
matrices rather than probability distributions.

{\em Definition of a significant subspace:}  Let $\Pi$ be the
projector onto a finite-dimensional vector space $V$.
For any density matrix
$\rho$ and $0 \leq \delta\leq 1$, we say that $V$ is a {\em significant
subspace} of $\rho$ (or a $\delta$-significant subspace) if
$\tr\Pi\rho \geq \delta$.

For any $\delta$ there can be many different significant subspaces and
the only upper bound we can place on their dimension is $\rank\rho$.
However, we will find it useful to examine the minimum dimension of
any $\delta$-significant subspace for a matrix $\rho$.  Denote this
dimension by $\sig(\rho,\delta)$ and define it by \be
\sig(\rho,\delta) = \left\{ \min \tr \Pi \big\arrowvert \Pi^2 = \Pi,
\tr \Pi \rho \geq \delta \right \} \ee

In discussing bipartite entanglement, the rank of a density matrix is
often useful to work with because it corresponds to the Schmidt number
of an entangled state, which can never increase under LOCC.
Unfortunately, in general a small perturbation can change the rank by
an arbitrary amount.  Significant subspaces are more stable under
perturbation, and we can use $\sig(\rho,\delta)$ to derive robust
bounds on the rank.
\begin{proposition}\label{prop:incompressibility}
If $0 \leq \delta\leq 1$ and $\rho,\sigma$ are density matrices with
$D(\rho,\sigma) \leq 2(1-\delta)$ then $\rank\sigma \geq
\sig(\rho,\delta)$
\end{proposition}
\begin{proof}
Let $P$ project onto the support of $\sigma$. Then $\tr P=\rank\sigma$
and $\tr P\sigma=1$.  Using Eq.~(\ref{eq:trace-dist}), we find
$2(1-\delta) \geq D(\rho,\sigma) \geq 2\tr
\Pi(\sigma-\rho)$ for any projector $\Pi$. Combining this with $\tr
P\sigma=1$, we obtain $\tr P\rho \geq \delta$.  From the definition
of $\sig$ we have that $\sig(\rho,\delta) \leq \tr P = \rank\sigma$.
\end{proof}

Significant subspaces are also preserved reasonably well under tensor
product.
\begin{proposition}\label{prop:sig-tensor}
Let $A$ and $B$ be density matrices with $\delta_A,\delta_B\geq 0$ and
$\delta_A + \delta_B \leq 1$.  Then
$\sig(A\otimes B, \delta_A + \delta_B) \geq
\sig(A\otimes B, \delta_A + \delta_B -\delta_A \delta_B ) > 
(\sig(A,\delta_A)-1) (\sig(B,\delta_B)-1)$.
\end{proposition}

{\em Remark:} Proposition~\ref{prop:sig-tensor} applies to
a density matrix that is a product of mixtures. Note that this
is a more restricted condition than the requirement that
the state is separable, which would have required only that
the state be a mixture of products. See \cite{jozsalinden} for
a discussion.

\begin{proof}
Let $a=\sig(A,\delta_A)-1$ and $b=\sig(B,\delta_B)-1$.
Define a projector $\Pi_A$ that projects onto the $a$ eigenvectors of
$A$ with the highest eigenvalues and likewise define $\Pi_B$ to
project onto the $b$ eigenvectors of $B$ with the highest eigenvalues.
From the definition of $\sig$, we know that $\tr \Pi_AA < \delta_A$
and $\tr  \Pi_B B < \delta_B$.

Now consider the orthogonal complements of $\Pi_A$ and $\Pi_B$.  Every
eigenvalue of $(\openone - \Pi_A)A$ has at least $a$ eigenvalues of
$A$ greater than or equal to it.  Likewise, every
eigenvalue of $(\openone - \Pi_B)B$ has at least $b$ eigenvalues of
$B$ greater than or equal to it.  Therefore, every
eigenvalue of 
\be\left((\openone - \Pi_A)\otimes(\openone - \Pi_B)\right)
(A\otimes B)\ee has at least $ab$ eigenvalues of
$A\otimes B$ greater than or equal to it.  Equivalently, the $ab$ highest
eigenvalues of $A\otimes B$ correspond to eigenvectors in the support
of $\openone - (\openone - \Pi_A)\otimes(\openone - \Pi_B)$.

Furthermore,
\bea
\tr (\openone - (\openone - \Pi_A)\otimes(\openone - \Pi_B))
(A\otimes B) &=& \\
 \tr \Pi_AA + \tr \Pi_BB - \tr\Pi_AA\tr\Pi_BB
&<& \delta_A + \delta_B - \delta_A \delta_B \n\eea

In the above, we make use of the fact that $\tr \Pi_AA < \delta_A$,
$\tr \Pi_B B < \delta_B$ and that $ \tr \Pi_AA + \tr \Pi_BB -
\tr\Pi_AA\tr\Pi_BB$ is an increasing function in both $\tr \Pi_AA$ and
$ \tr \Pi_B B$.
 
Thus, the largest $ab$ eigenvectors of $A\otimes B$ have weight less
than $\delta_A + \delta_B- \delta_A \delta_B $.  This implies
\be 
ab < 
\sig(A\otimes B,\delta_A + \delta_B - \delta_A \delta_B )
\leq \sig(A\otimes B, \delta_A + \delta_B),\ee
the desired result.
\end{proof}

One application of typical subspaces is to show that
$\rho^n$ can be compressed to $2^{nS(\rho)+O(\sqrt{n})}$
dimensions with asymptotically small error.  This is because for large
$n$, the spectrum of $\rho^n$ (for any $\rho$) approaches a Gaussian
distribution and almost all eigenvalues are between
$2^{-nE-O(\sqrt{n})}$ and $2^{-nE+O(\sqrt{n})}$ \cite{CT}. To prove
that improving upon this result is impossible, we will examine the
significant subspaces of $\rho^n$.

First we will need to state central limit theorem in a manner that
bounds the rate of convergence to Gaussianity.
\begin{lemma}[Berry-Esse\'en]\label{lemma:clt}
Let $\rho$ have eigenvalues $p_1,\ldots,p_d$ and define $E=-\sum_i
p_i\log p_i$, $\alpha^2 = \sum_i p_i(\log p_i + E)^2$ (with
$\alpha>0$) and $\beta=\sum_i p_i|\log p_i + E|^3$.  Let $\mu(a,b)$
denote the sum of all eigenvalues of $\rho^n$ between $2^a$ and $2^b$.
Then, for all $a\leq b$, \be \left|\mu(a,b) -
N\left(\frac{a+nE}{\sqrt{n}\alpha},\frac{b+nE}{\sqrt{n}\alpha}\right)
\right | < \frac{25\beta}{\sqrt{n}}
\label{eq:clt-discrepancy}\ee
where $N(x_1,x_2)$ is the cumulative normal distribution
\be
N(x_1,x_2) \equiv \frac{1}{\sqrt{2\pi}} \int_{x_1}^{x_2}
e^{-\frac{x^2}{2}} dx\ee
\end{lemma}

\begin{proof}
A proof can be found in \cite{Bolt84}.
\end{proof}

As a corollary, for any $\rho$ and $\delta$ there exists $n_0$ such
that for $n\geq n_0$ the left side of Eq.~(\ref{eq:clt-discrepancy}) is
less than $\delta$.  Another useful consequence is that $n$ copies of
a state with entropy $E$ have significant subspaces of dimension
$2^{nE\pm O(\sqrt{n})}$.  

\begin{proposition}\label{prop:binomial-significant}
Let $\rho, E, \alpha$ and $\beta$ be as in Lemma~\ref{lemma:clt}.
Then there exist $\delta<1, C, n_0$ such that 
$\sig(\rho^n,\delta) > C2^{nE+\alpha\sqrt{n}}$ for all $n\geq n_0$.
\end{proposition}

Here $\delta$ and $C$ are universal constants and $n_0$ depends only
on $\beta$; one valid choice would be $\delta=0.95$, $C=0.01$ and
$n_0=10^7\beta^2$.

\begin{proof}
Let $P$ be the projector onto the eigenvectors of $\rho^n$ with
eigenvalues above $2^{-nE-1.1\alpha\sqrt{n}}$.  Applying
Lemma~\ref{lemma:clt} yields $\tr P\rho^n \leq N(-1.1,\infty) +
\frac{25\beta}{\sqrt{n}} \approx 
0.94 + \frac{25\beta}{\sqrt{n}}$, which for some choice
of $n_0$ and $\delta$ can be guaranteed to be less than
$\delta$. Since $P$ picks out the largest eigenvectors, it
minimizes the dimension of a $\delta$-significant subspace.
Therefore, we have $\sig(\rho^n,\delta) > \tr P$.  Now, to lower-bound $\tr
P$, note that the weight of eigenvalues of $\rho^n$ between
$2^{-nE-1.1\alpha\sqrt{n}}$ and $2^{-nE-\alpha\sqrt{n}}$ is at least
$N(-1.1,-1)-\frac{25\beta}{\sqrt{n}}$
which we can 
make greater than some constant $C>0$.  Since the eigenvalues in
this region are no greater than $2^{-nE-\alpha\sqrt{n}}$ it follows
that $C 2^{nE+\alpha\sqrt{n}} \leq \tr P\leq \sig(\rho^n,\delta)$.
\end{proof}

{\em Remark:} Proposition~\ref{prop:binomial-significant} can
be generalized to show that for any $r >0$,
there exist $\delta_r < 1$, $C$ and $n_0$
such that $\sig(\rho^n,\delta_r) >
C2^{nE+ r \alpha\sqrt{n}}$ for all $n\geq n_0$.

We are now ready to prove our main result.

\section{Proof of the main theorem}\label{sec:main-proof}
\subsection{The inefficiency bound}
Combining Proposition~\ref{prop:incompressibility} with
Proposition~\ref{prop:binomial-significant}, we find that any state
$\sigma$ with $D(\sigma,\rho^n)<2(1-\delta)$ must have support on a
space of dimension $\Omega(2^{nE+\alpha\sqrt{n}})$
 \footnote{This means that to guarantee $O(1)$ error, Schumacher
compression (and hence the dilution protocol given by BBPS) must have
$\Omega(\sqrt{n})$ inefficiency.  We can use a similar argument to
prove that this inefficiency bound applies to {\em any} dilution
protocol.}.
This result allows us to prove that the inefficiency (loss
of entanglement) bound applies to {\em any} dilution protocol.

\begin{proof}[of part a) of Theorem~\ref{thm:main-result}]  Here we
will not need the protocol reduction of
Section~\ref{sec:protocol-reduction}.  Instead, represent a general
quantum operation as a map from pure states to ensembles of pure
states.  For example, the $\$$ operation of
Section~\ref{sec:protocol-reduction} can be said to map the pure state
$a\ket{0}+b\ket{1}$ to the ensemble $\{p_i,\ket{i}\}_{i=0,1}$ with
$p_0=|a|^2$ and $p_1=|b|^2$.  Now suppose that Alice and Bob start
with $\ket{\Phi_d}$, obtain a series of measurement outcomes that
indicate the protocol has succeeded, and end with an ensemble of
states $\{p_i, \ket{\varphi_i}\}$.  By ``success'' we mean that their
resulting density matrix is close to the desired state; i.e.
$D(\sum_i p_i \ket{\varphi_i}\bra{\varphi_i},\ket{\psi}^n)<\epsilon$.

Since the Schmidt number never has any chance of increasing, we must
have $\mbox{Sch}\,\ket{\varphi_i}\leq d$ for all $i$.  Furthermore,
since the trace distance is convex, there exists an $i$ for which
$D(\ket{\varphi_i},\ket{\psi}^n)<\epsilon$.  Let $\rho=\tr_B
\ket{\psi}\bra{\psi}$, let $E, \alpha$ and $\beta$ be as in
Lemma~\ref{lemma:clt} and choose $\delta$, $C$ and $n_0$ according to
Proposition~\ref{prop:binomial-significant}.  Then choose $\epsilon_0$
such that $\epsilon_0<2(1-\delta)$.  Since tracing out Bob's system
cannot increase the trace distance, we have
$D(\tr_B\ket{\varphi_i}\bra{\varphi_i}, \rho^n) < \epsilon <
2(1-\delta)$.  Again we apply Propositions
\ref{prop:incompressibility} and \ref{prop:binomial-significant} to
find that \be C 2^{nE+\alpha\sqrt{n}} \leq
\rank\tr_B\ket{\varphi_i}\bra{\varphi_i} = \mbox{Sch}\,\ket{\varphi_i}
\leq d \ee proving the desired result (up to an overall constant that
we can absorb into $\alpha$ and $n_0$).
\end{proof}

\subsection{The communication bound}
In section~\ref{sec:protocol-reduction}, we reduced an arbitrary
dilution protocol to one consisting of a generalized measurement by
Alice, a local unitary by Bob conditioned on the result, and then the
discarding of ancilla systems by both parties.

Alice's measurement is the interesting step, since we can relate the
spectrum of the measurement operator to the Schmidt
coefficients of the resulting state.  If the final state roughly
resembles $\ket{\psi}^n$, then the measurement outcome that produced
it must also have a typical subspace (say of weight 0.99) with
dimension $2^{\Omega(\sqrt{n})}$ times larger than some significant
subspace (of weight around 0.25).  We will show that any such
measurement outcome occurs with probability $2^{-\Omega(\sqrt{n})}$
and thus that $\Omega(\sqrt{n})$ bits of communication are
necessary to tell Bob the result of Alice's measurement.

To deal with the complication of discarding ancillas, we state one final
lemma. 

\begin{lemma}\label{lemma:mixed-pure}
Consider a bipartite Hilbert space $\H_{AB}$ with states
$\ket{\psi}\in\H_{AB}, \ket{\phi}\in\H_A$ and \be
D\left(\tr_B\ket{\psi}\bra{\psi},\ket{\phi}\bra{\phi}\right) <\epsilon
\label{eq:mixed-pure-hypothesis}\ee
Then $\exists\ket{\gamma}\in\H_B$ such that
$$D\left(\ket{\psi}, \ket{\phi} \otimes \ket{\gamma} \right) <
2\epsilon$$
\end{lemma}

In other words, if tracing out $\H_B$ leaves the state almost pure,
then the subsystems must have been almost separable to begin with.
This is proved in Appendix~\ref{appendix:mixed-pure}.

Now we can proceed with the proof of the classical communication bound.

\begin{proof}[of part b) of Theorem \ref{thm:main-result}]
By Claim~\ref{claim:protocol-reduction}, it suffices to consider an
entanglement dilution procedure with the following form.

\begin{figure}[htbp]
\includegraphics[keepaspectratio,width=3.5in]{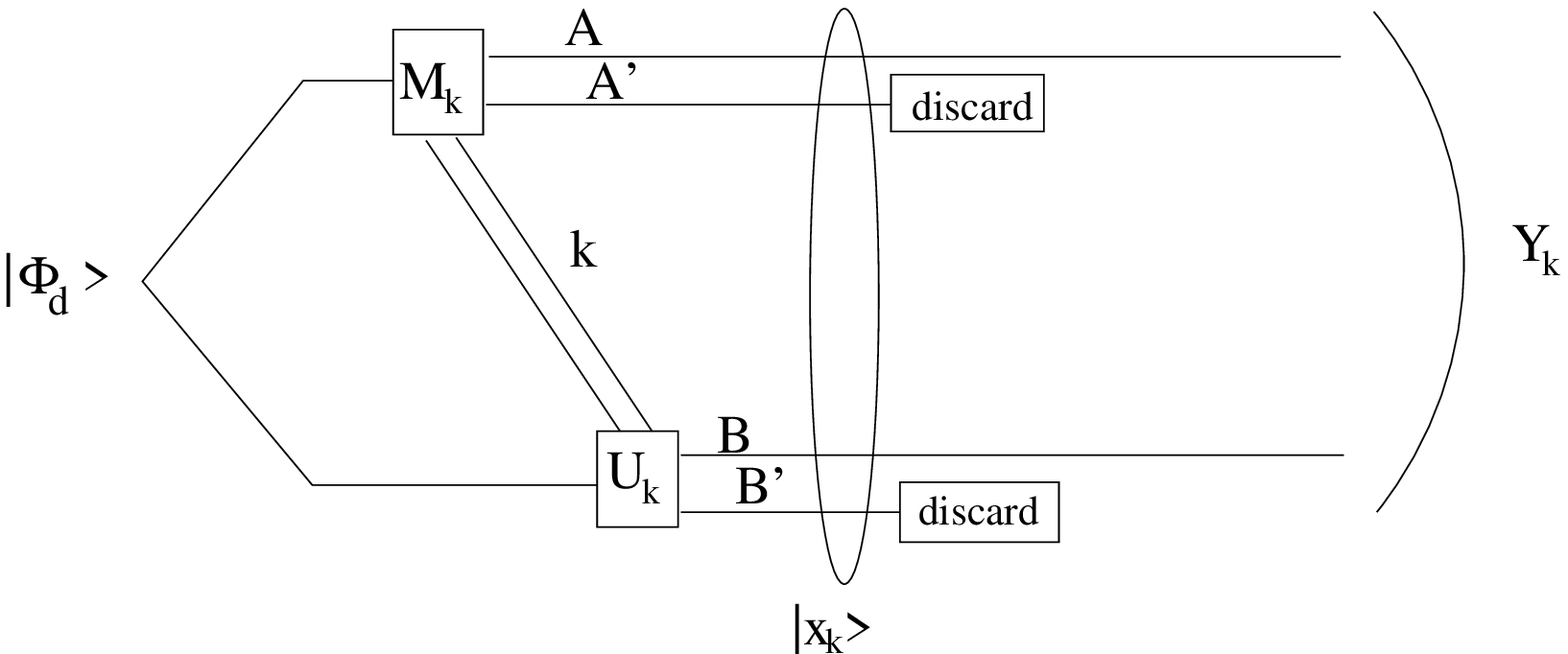}
\end{figure}

Alice first performs a generalized measurement $\{M_k\}$ with $2^c$
outcomes on her half of $\ket{\Phi_d}$ and transmits her $c$-bit
measurement outcome $k$ to Bob, who performs some unitary operation
$U_k$ conditioned on the result.  The result of the measurement is $k$
with probability $\frac{1}{d}\tr M_kM_k^\dag$, leaving Alice and Bob
with the pure state $\ket{x_k}_{AA'BB'}$, which can be written as
\be\left(M_k\otimes U_k\right)\ket{\Phi_d} = \sqrt{\frac{\tr
M_kM_k^\dag}{d}}\ket{x_k} \label{eq:povm} \ee
Then Alice and Bob trace out subsystems
$A'$ and $B'$, leaving the state $Y_k=\tr_{A'B'}\ket{x_k}\bra{x_k}$.
When the protocol succeeds (which occurs with probability $\geq
2^{-s}$), then $Y_k$ should be close to the desired state
$\ket{\psi}^n$.  Specifically, we should have
\be D\left(Y_k,\ket{\psi}^n\right) < \epsilon
\label{eq:success-condition}\ee

Since there are at most $2^c$ successful values of $k$
(i.e. measurement outcomes for which Eq.~(\ref{eq:success-condition})
holds), there must exist at least one successful value of $k$ that has
probability $\geq 2^{-(c+s)}$ of occuring.  Equivalently,
\be 
\tr M_kM_k^\dag \geq \frac{d}{2^{c+s}}
\label{eq:prob-condition}\ee

We will use this value of $k$ implicitly for the rest of the proof,
and refer to $M$, $U$, $\ket{x}$, $Y$ and so forth instead of $M_k$,
$U_k$, $\ket{x_k}$ and $Y_k$.

Now $D(Y,\ket{\psi}^n)=D(\tr_{A'B'}\ket{x}\bra{x},
\ket{\psi}^n)<\epsilon$.  Thus we can use
Lemma~\ref{lemma:mixed-pure} to show that
$\exists\ket{\gamma}\in\H_{A'B'}$ such that
\be D(\ket{x},\ket{\psi}^n\ket{\gamma})<2\epsilon
\label{eq:x-almost-separable} \ee

Define $\rho=\tr_B\ket{\psi}\bra{\psi}$, $X=\tr_{BB'}\ket{x}\bra{x}$
and $\Gamma=\tr_{B'}\ket{\gamma}\bra{\gamma}$.  Since tracing out a
subsystem never increases distance between two states
\be D(X,\rho^n\otimes\Gamma)<2\epsilon 
\label{eq:X-almost-separable}\ee

At this point, several different variables have been introduced to
label different subsystems of $\ket{x}$.  To keep track of their
relations to one another, the following diagram may be useful.

$$\begin{CD}
\ket{x_k}\approx \ket{\psi}^n\otimes\ket{\gamma}
@>{\tr_{A'B'}}>>
Y_k \approx \ket{\psi}^n \\
@VV{\tr_{BB'}}V   @VV{\tr_B}V \\
X \approx \rho^n\otimes\Gamma
@>{\tr_{A'}}>>
\tr_{A'}X \approx \rho^n
\end{CD}$$

Alice's reduced density matrix, $X$, turns out to have a simple
expression in terms of $M$ that will make it quite useful to work
with.  From Eq.~(\ref{eq:povm}),

\begin{eqnarray}
{ Tr MM^\dag \over d}  \ket{x}\bra{x} & =&
 ( M \otimes  U ) | \Phi_d \rangle \langle \Phi_d |
( M^\dag \otimes  U^\dag) \nonumber \\
{ Tr MM^\dag \over d} Tr_{BB'}  | x \rangle \langle x | & =&
M ( { I_d \over d}) M^\dag 
=  ( { 1 \over d}) M M^\dag \nonumber \\
 X = Tr_{BB'}  | x \rangle \langle x | & =& { M M^\dag \over  Tr MM^\dag } .
\end{eqnarray}
Thus $\|X\| =
\frac{\|MM^\dag\|}{\tr MM^\dag} \leq \frac{1}{\tr MM^\dag}$.  Plugging
in Eq.~(\ref{eq:prob-condition}) gives
\be \|X\| \leq\frac{2^{c+s}}{d}\label{eq:X-norm-CC}\ee
 Since Schmidt number can never increase by local operations, we also
have that
\be\rank X\leq d \label{eq:X-rank-bound}\ee

The proof now follows from Eqs.~(\ref{eq:X-almost-separable}),
(\ref{eq:X-norm-CC}) and
(\ref{eq:X-rank-bound}) and our results about significant subspaces.

First define $P_1\in\B(\H_A)$ to be the projector onto all
eigenvectors of $\rho^n$ with eigenvalue of $2^{-nE}$ or greater.
There can be no more than $2^{nE}$ such eigenvectors, so $\tr P_1\leq
2^{nE}$.  From Lemma~\ref{lemma:clt}, we have that
\be \tr P_1\rho^n \geq \frac{1}{2} - \frac{25\beta}{\sqrt{n}} >
\frac{1}{4}\ee
where the last inequality holds as long as $n_0 > 2500\beta^2$.  In
terms of significant subspaces, we can summarize this with \be
\sig(\rho^n,\frac{1}{4}) < \tr P_1 \leq 2^{nE}
\label{eq:rho-sig-trace}\ee

We will now use our bound on the rank of $X$ from
Eq.~(\ref{eq:X-rank-bound}) to show that $\Gamma$ also has a small
significant subspace.
To accomplish this we will seek constants $\delta_\rho$ and
$\delta_\Gamma$ with the properties:
\begin{itemize}
\item $\delta_\rho + \delta_\Gamma + \epsilon < 1$
\item $\delta_\Gamma > 4\epsilon_0$
\item 
$\forall n\geq n_0, \sig(\rho^n,\delta_\rho) >
C2^{nE+\alpha\sqrt{n}}$
 where $n_0$ depends on $\rho$ and $C$ does
not.
\end{itemize}
According to Proposition~\ref{prop:binomial-significant},
this last condition is met by $\delta_\rho=0.95$.  To
meet the other two, it will suffice to set
$\epsilon \leq \epsilon_0=0.01$ and $\delta_\Gamma=0.04$.

Combining $\delta_\rho + \delta_\Gamma + \epsilon < 1$
with Eq.~(\ref{eq:X-almost-separable}) and
Proposition~\ref{prop:incompressibility} yields
\ben \sig(\rho^n\otimes\Gamma,\delta_\rho + \delta_\Gamma) &<&
\sig(\rho^n\otimes\Gamma,1-\epsilon)\\
&\leq& \rank X \leq d
\een

Applying Proposition~\ref{prop:sig-tensor} now gives
\bea
\sig(\Gamma,\delta_\Gamma) &\leq&
\frac{\sig(\rho^n\otimes\Gamma,\delta_\rho + \delta_\Gamma)}
{\sig(\rho^n,\delta_\rho) - 1} + 1\n\\
&<& \frac{d}{C2^{nE+\alpha\sqrt{n}}-1} + 1 \n\\
&\approx& \frac{d}{C2^{nE+\alpha\sqrt{n}}}
\label{eq:gamma-sig-trace}
\eea
In the last line, the factors of $+1$ and $-1$ are negligible compared
with the exponentials in the numerator and denominator, so we can
absorb them into $C$.

Define $P_2$ to project onto the highest $\sig(\Gamma,\delta_\Gamma)$
eigenvalues of $\Gamma$.  Then $\tr P_2\Gamma\geq \delta_\Gamma$ and
$\tr P_2 =\sig(\Gamma,\delta_\Gamma) \leq
\frac{d}{C2^{nE+\alpha\sqrt{n}}}$.

Now we combine Eqs.~(\ref{eq:X-norm-CC}), (\ref{eq:rho-sig-trace})
and (\ref{eq:gamma-sig-trace}) to obtain
\bea
\tr (P_1 \otimes P_2)X &\leq&
\tr P_1 \tr P_2 \|X\| \n\\
&\leq&
2^{nE}\cdot
\frac{d}{C2^{nE+\alpha\sqrt{n}}}\cdot
\frac{2^{c+s}}{d}
\n\\
&=& \frac{2^{c+s}}{C2^{\alpha\sqrt{n}}}
\eea

On the other hand, $P_1$ and $P_2$ project onto significant subspaces
of $\rho^n$ and $\Gamma$ respectively, so
\be \tr (P_1 \otimes P_2) (\rho^n \otimes \Gamma) \geq
\frac{1}{4}\delta_\Gamma \ee

Thus
 \ben 2\epsilon_0 &\geq& 2\epsilon \geq
D(\rho^n\otimes\Gamma,X) \\ 
&\geq& 2\tr (P_1\otimes P_2) (\rho^n\otimes\Gamma-X)\\ 
&\geq& 
2\left(\frac{\delta_\Gamma}{4}
 - C^{-1}2^{c+s-\alpha\sqrt{n}}
 \right) \een 

Solving for $c+s$ yields 
\be c+s \geq \alpha\sqrt{n} + 
\log\left(\frac{\delta_\Gamma}{4} - \epsilon_0\right)
-\log C\ee
 So there exist choices of
$\alpha,n_0,\epsilon_0$ that prove the theorem.

\end{proof}

{\em Remark:} This bound only assumes $O(1)$ error; specifically
$\epsilon_0=0.01$.  That is to say, even an entanglement dilution protocol
with a non-negligible amount of loss of fidelity is still covered by
Theorem~\ref{thm:main-result}, which is, therefore, a rather strong result.

{\em Remark:} Suppose we are interested in
$o(1)$ error, a stricter requirement.  It is not
difficult to improve our classical communication
bound to $\omega(\sqrt{n})$; by analogy with data compression,
achieving error $\epsilon$ can be shown to require inefficiency and
classical communication of $\Omega(\alpha\sqrt{n\log(1/\epsilon)})$.
See the Remark just after Proposition \ref{prop:binomial-significant}
for the main modification needed in the proof.

\section{Conclusions and Discussions}\label{sec:conclusion}
We have proven that entanglement dilution from any number of singlets
to $n$ pairs of bipartite partially entangled pure states necessarily
requires $\Omega(\sqrt{n})$ bits of classical communication, thus
showing that the main result in \cite{LP99} is, in fact,
optimal.  

A technique of our proof is a generalization of the
reduction result of \cite{LP97} in entanglement manipulation protocols.
\cite{LP97} showed that any exact pure state LOCC entanglement manipulation
protocol could have two-way communication reduced to one-way
communication.  We extended their result to protocols ending in mixed
states, and proved that Alice can be assumed to transmit her entire
measurement outcome with no increase in communication.  This reduction
applies rather generally and may be useful in deriving bounds on
classical communication for other sorts of entanglement manipulations.

An interesting, but difficult, problem is to try to derive similar
bounds for transformations from a single copy of one state
$\ket{\psi_1}$ to a single copy of another, $\ket{\psi_2}$.  Here,
there are no intermediate results between the constant lower bound of
\cite{LP97} and the construction of \cite{JS01} and \cite{Lo00} which
uses $\log\rank\tr_B\ket{\psi_1}\bra{\psi_1}$ bits of communication.

A more specific, and perhaps more tractable, problem is that of
converting between many copies of different partially entangled
states.  Suppose we constrain these sorts of interconversions to waste
only $o(n)$ bits of entanglement.  Recall from \cite{LP99} that
dilution can be performed with $O(\sqrt{n})$ bits of classical
communication (and that \cite{Lo00} shows how to reduce this cost by a
factor of 2).  Our lower bound in Theorem~\ref{thm:main-result}
matches this construction up to a constant factor.  If we could
improve either our lower bound or the protocol of \cite{LP99,Lo00} to
eliminate this constant factor, then we could prove that, at least for
some partially entangled states $\ket{\psi_1}$ and $\ket{\psi_2}$,
preparing $\ket{\psi_2}^n$ from
$\ket{\psi_1}^{n\frac{E(\psi_2)}{E(\psi_1)}+o(n)}$ requires
$\Omega(\sqrt{n})$ bits of communication.

To see this, consider the following two processes.  Process A:
Start from singlets, dilute to $n$ copies of $\ket{\psi_2}$ directly.
Process B: Start from singlets, dilute first to
$n\frac{E(\psi_2)}{E(\psi_1)}+o(n)$ copies of $\ket{\psi_1}$ and then
apply a conversion procedure from the
$n\frac{E(\psi_2)}{E(\psi_1)}+o(n)$ copies of $\ket{\psi_1}$ to $n$
copies of $\ket{\psi_2}$.  From Theorem~\ref{thm:main-result}, Process
A takes $\Omega(\alpha_{\psi_2} \sqrt{n})$ bits of classical
communication. Process B is a way of realizing Process A and is, thus,
constrained by our lower bound.  Therefore, it must also take
$\Omega(\alpha_{\psi_2} \sqrt{n})$ bits of classical communication.
Now, \cite{LP99, Lo00} showed that Part~1 of Process B can be done
with only $O\left(\alpha_{\psi_1}
\sqrt{n\frac{E(\psi_2)}{E(\psi_1)}}\right)$ bits of classical
communication.  If the constant factors of
Theorem~\ref{thm:main-result} and \cite{LP99, Lo00} were the same then
the classical communication cost of Part~2 of Process~B would be
lower-bounded by an amount of that is at least the difference between
these two numbers, which is positive whenever
\be\frac{\alpha_{\psi_2}}{\sqrt{E(\psi_2)}} >
\frac{\alpha_{\psi_1}}{\sqrt{E(\psi_1)}}.
\label{eq:partial-CC-cond}\ee
This would establish a total ordering on entangled states.  If,
instead, we were unable to close the gap between the constants of the
upper and lower bounds, then we would have a partial ordering; for
some constant $C>1$, converting $\ket{\psi_1}$ to $\ket{\psi_2}$ would
require communication whenever
$\frac{\alpha_{\psi_2}}{\sqrt{E(\psi_2)}} > C
\frac{\alpha_{\psi_1}}{\sqrt{E(\psi_1)}}$.

  It is an open question whether Eq.~(\ref{eq:partial-CC-cond}) is a
necessary condition for a classical communication bound or whether
such a result holds for more general pairs of partially entangled
states.  Also, unlike the case of diluting from maximally entangled
states, there may be a tradeoff between inefficiency and communication
when starting with partially entangled states.

We can also apply our main theorem to the resource model of quantum
information processing, where it implies that there is a limited
extent to which the resource of entanglement can be thought of as
independant of form.  Thus, when one considers scenarios with
prior shared entanglement, one should
either a) restrict the scenario to $o(n^2)$ copies of any partially
entangled state (if the protocol calls for $O(n)$ bits of classical
communication) or b) specify explicitly what forms of bipartite
entanglement are allowed.

More generally, the classical communication cost in quantum
information processing is an important subject \cite{Lo00}. One rather
curious fact about our result is that dilution requires a large amount
of classical communication, but there does not appear to be any simple
way to use it for signaling.  It would be interesting to determine
whether any black box capable of performing entanglement dilution
could also be used to transmit information, as this would provide an
intuitive alternate proof of our main result.

Two final remarks are in order. First, the classical
bits transmitted in entanglement dilution constitute classical shared
randomness between Alice and Bob. Such classical shared randomness can
be an important resource in information processing.  However since
those bits are sent through a classical channel, their value is
potentially public knowledge.  Thus, we can call them {\em shared
public randomness}.  Second, the loss of entanglement in, for example,
entanglement concentration will generally give rise to shared
randomness that is private to Alice and Bob.  Similarly, the loss of
entanglement in entanglement dilution will give rise to residual
correlations that are almost uncorrelated to the desired final state
(according to Eq.~(\ref{eq:x-almost-separable})).  Such randomness is
private to Alice and Bob, and could therefore be used for applications
such as a cryptographic one-time-pad \footnote{However, Alice and Bob
also have the option of keeping the ancillary state quantum, rather
than measuring it and generating classical randomness.  According to
Lemma~\ref{lemma:mixed-pure}, this ancillary state is nearly separable
from the output of the dilution procedure.  Thus, Alice and Bob could
consider keeping their ancillas and later recycling some of its
entanglement.}.  We believe that a complete theory of the resource
model of quantum information processing should take full account of
these two resources---shared public randomness and shared private
randomness.

After the completion and circulation of a draft version of the current
paper, we became aware of the independent proof of almost the same
result but with a different approach by Patrick Hayden and Andreas
Winter \cite{HW02}.

\begin{acknowledgements}
We thank helpful discussions with colleagues including Charlie
Bennett, Andrew Childs, Isaac Chuang, Debbie Leung, Sandu Popescu, Ben
Recht, John Smolin and Jason Taylor.  We are also indebted to Patrick
Hayden and Andreas Winter for sharing their draft of \cite{HW02} with
us and for enlightening discussions.  AWH was supported in part
by the National Security Agency (NSA) and Advanced Research and
Development Activity (ARDA) under Army Research Office (ARO) contract
number DAAD19-01-1-06.
\end{acknowledgements}

\appendix
\section{Proof of Lemma~\ref{lemma:mixed-pure}}
\label{appendix:mixed-pure}

\begin{proof}
For any density matrices $\rho_0$ and $\rho_1$, Eq.~(46) of \cite{FG97}
states that \be 1 - F(\rho_0,\rho_1) \leq
\frac{1}{2}\tr|\rho_0-\rho_1|
\label{eq:trace-fidelity}\ee
where $F(\rho_0,\rho_1)=\tr\sqrt{\sqrt{\rho_0}\rho_1\sqrt{\rho_0}}$ is
the fidelity.  By Uhlmann's theorem \cite{Uhl76} \be F(\rho_0,\rho_1)
= \max_{\varphi_0,\varphi_1} |\braket{\varphi_0}{\varphi_1}|\ee where
$\varphi_0$ and 
$\varphi_1$ are purifications of $\rho_0$ and $\rho_1$, respectively.
Equivalently we can fix an arbitrary purification $\varphi_0$ and
maximize only over $\varphi_1$.

 Applying these two results we find that there exists
$\ket{\gamma}\in\H_B$ such that
\be|\bra{\psi}(\ket{\phi}\otimes\ket{\gamma})|\geq 1-\epsilon/2\ee
since we can consider $\ket{\phi}\otimes\ket{\gamma}$ to be a
purification of $\ket{\phi}\bra{\phi}$.

To obtain the trace distance between these states, write
$\ket{\phi}\otimes\ket{\gamma}$ as $a\ket{\psi}+b\ket{\psi^\perp}$,
where $|a|\geq 1-\epsilon/2$ and
$|b|=\sqrt{1-|a|^2}\leq\sqrt{\epsilon-\epsilon^2/4}$.  The trace
distance is then given by \be\tr\left| \left(\begin{array}{cc}1-|a|^2
& 0 \\ 0 & -|b|^2\end{array}\right) \right| = 2 |b|^2 \leq
2\epsilon-\frac{\epsilon^2}{2} < 2\epsilon\ee
\end{proof}

\end{document}